\documentclass[times,twocolumn,final]{elsarticle}

\usepackage{medima}
\usepackage{framed,multirow}

\usepackage{amssymb}
\usepackage{latexsym}

\usepackage{url}

\usepackage{hyperref}

\newcommand{\tabincell}[2]{\begin{tabular}{@{}#1@{}}#2\end{tabular}} 
\usepackage{textcomp}
\usepackage{threeparttable}
\usepackage{amsmath}
\usepackage{multirow}
\usepackage{gensymb}
\usepackage{graphicx}
\usepackage{booktabs} 
\usepackage{float}
\usepackage{color}
\usepackage[noend]{algpseudocode}
\usepackage{algorithmicx,algorithm}
\usepackage{times}
\usepackage{epsfig}
\usepackage{enumerate}
\usepackage{verbatim}
\usepackage{cite}
\usepackage{amsfonts}
\usepackage{array}
\usepackage{comment}
\usepackage{multicol}
\usepackage{multirow}
\usepackage{epsfig}
\usepackage{booktabs}
\usepackage{float}
\usepackage{graphicx}
\usepackage{amsmath,bm}
\usepackage{amssymb}
\usepackage{amsfonts}
\usepackage{amsthm}
\usepackage{graphicx,lipsum}
\usepackage{url}

\definecolor{newcolor}{rgb}{.8,.349,.1}

\journal{Medical Image Analysis}

\begin{document}

\verso{W. Nie \textit{et~al.}}

\begin{frontmatter}
\title{Deep Reinforcement Learning Framework for Thoracic Diseases Classification via Prior Knowledge Guidance}

\author[1]{Weizhi Nie}
\author[1]{Chen Zhang}
\author[1]{Dan Song\corref{cor1}}
\cortext[cor1]{Corresponding author}
\ead{dan.song@tju.edu.cn}
\author[2]{Lina Zhao}
\author[5,6]{Yunpeng Bai}
\ead{oliverwhite@126.com}
\author[2,3,4]{Keliang Xie}
\author[1]{Anan Liu}

\address[1]{School of Electrical and Information Engineering, Tianjin University, Tianjin 300072, China}
\address[2]{Department of Critical Care Medicine, Tianjin Medical University General Hospital, Tianjin 300052, China}
\address[3]{Department of Anesthesiology, Tianjin Medical University General Hospital, Tianjin 300052, China}
\address[4]{Tianjin Institute of Anesthesiology, Tianjin 300052, China}
\address[5]{Department of Cardiac Surgery, Chest Hospital, Tianjin University, Tianjin 300222, China}
\address[6]{Clinical school of Thoracic, Tianjin Medical University, Tianjin 300052, China}

\received{1 May 2013}
\finalform{10 May 2013}
\accepted{13 May 2013}
\availableonline{15 May 2013}
\communicated{S. Sarkar}

\begin{abstract}
The chest X-ray is often utilized for diagnosing common thoracic diseases. In recent years, many approaches have been proposed to handle the problem of automatic diagnosis based on chest X-rays. However, the scarcity of labeled data for related diseases still poses a huge challenge to an accurate diagnosis. In this paper, we focus on the thorax disease diagnostic problem and propose a novel deep reinforcement learning framework, which introduces prior knowledge to direct the learning of diagnostic agents and the model parameters can also be continuously updated as the data increases, like a person's learning process. Especially, 1) prior knowledge can be learned from the pre-trained model based on old data or other domains' similar data, which can effectively reduce the dependence on target domain data, and 2) the framework of reinforcement learning can make the diagnostic agent as exploratory as a human being and improve the accuracy of diagnosis through continuous exploration. The method can also effectively solve the model learning problem in the case of few-shot data and improve the generalization ability of the model. Finally, our approach's performance was demonstrated using the well-known NIH ChestX-ray 14 and CheXpert datasets, and we achieved competitive results. The source code can be found here: \url{https://github.com/NeaseZ/MARL}. 
\end{abstract}


\end{frontmatter}







\section{Introduction}
\label{sec:introduction}
\begin{figure}
\centering
\includegraphics[width=0.95\linewidth]{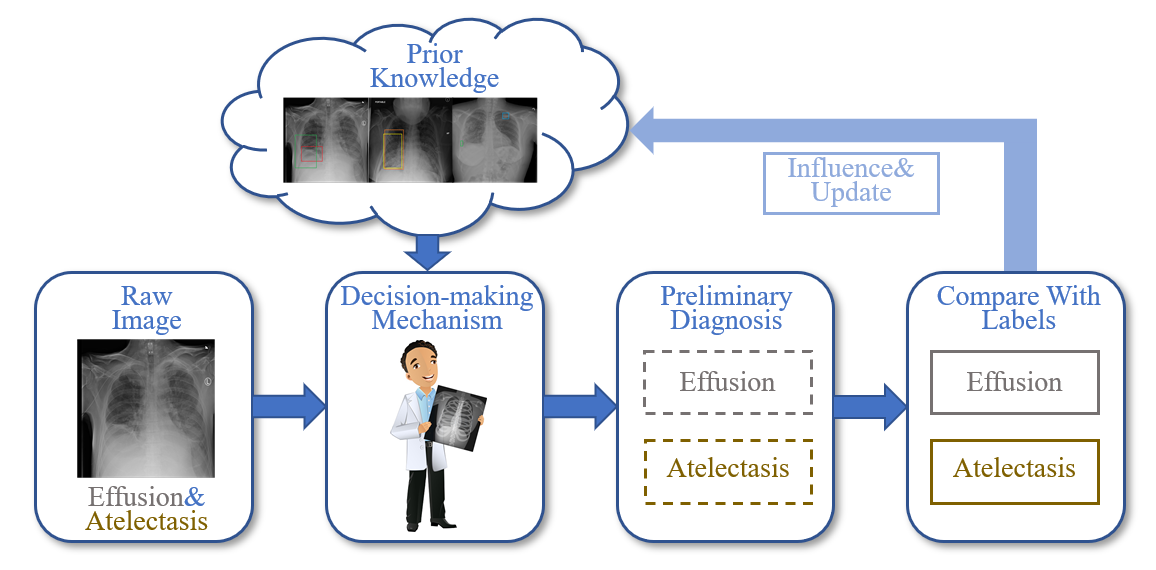}
\caption{The motivation of our approach: Each diagnosis can be based on the doctor's experience which can be seen as the utilization of the doctor's prior knowledge. Meanwhile, the diagnosis can also increase or update the experience of the doctor, which then helps to make a more accurate diagnosis in the future.}
\label{motivation}
\end{figure}
The chest X-ray (CXR) is one of the most commonly used clinical examinations in clinical scenarios. Early screening of many diseases relies on CXR data, such as heart disease, lung infection, throat examination, etc. This is because CXR data is easier to obtain and costs less. Traditional diagnosis and treatment rely heavily on the experience of full-time doctors. Effective automated diagnosis and treatment technology can greatly improve diagnosis and treatment efficiency while also assisting doctors in improving the accuracy of early patient diagnosis and treatment. However, the sparsity of medical data and the diversity of patients' clinical outcomes have been major challenges in the development of automated diagnosis and treatment technology.

In recent years, many proposed approaches, like \citet{fu2021tortuosity,abbas2021classification} have handled the CXR image diagnosis problem as a multi-label classification problem. Their work relies on the large scale of training data and parameter selection. These approaches are difficult to improve further because of the high inner similarity\citet{sarvamangala2021convolutional} between classes in CXR, their traditional convolutional network-based methods cannot distinguish well. 
Furthermore, many medical image datasets are difficult and expensive to achieve on a large scale, so improving data use efficiency is critical.
In recent years, some researchers have focused on how to achieve high classification accuracy based on the few-shot learning technique. 
Cherti.\textit{et al.}\citet{cherti2021effect} combined pre-training and transfer learning to large-scale natural and medical images datasets and verify the effectiveness of their method by some experiments. 
Singh.\textit{et al.}\citet{singh2021metamed} formulated a few-shot learning problem and presented a meta-learning-based model, which can adapt to rare disease classes with the availability of a few images. 
Furthermore, some methods consider utilizing external knowledge as auxiliary information to improve the performance of the model.
Chen\textit{et al.}\citet{chen2022boosting} proposed a multi-label annotation framework based on a external medical knowledge graph and model relevant lesion labels more comprehensively, but the professional medical knowledge graph is always hard to obtain and difficult to generalize to other tasks.

\subsection{Motivation}
Based on the above analysis, we hope the automatic diagnosis model could consider the context of a medical image and also utilize some additional prior knowledge like that of a professor doctor. Meanwhile, we also hope the automatic diagnosis model can be updated like the learning process of a doctor based on the increase in clinic data, like Fig.\ref{motivation}.  
Thus, the learning process can be divided into two parts: 1) Obtain experience and update: the experience can be obtained from the other domain or the target domain data. The experience can be seen as the pre-trained module's output.
2) Make diagnosis: The module makes the final diagnosis based on the current medical image and also according to the prior knowledge learned from old data or other domain's data.

Here, each diagnosis is influenced by the previous experience (previous module's parameter), and each diagnosis can also increase or update the experience, which is very similar to the human's knowledge growing process. Obviously, this process can be seen as a Markov decision process (MDP). Thus, it is natural that the reinforcement learning (RL) framework can be used to represent or estimate this process. However, we also need to solve two questions below:
\begin{itemize}
    \item How to learn and save the prior knowledge based on limited image data. The prior knowledge is a very abstract theory, so we need to define the format of the prior knowledge, which should be represented and utilized for helping to make the diagnosis, then ease the problem of catastrophic forgetting.
    \item How to utilize the prior knowledge to guide the training of the diagnosis model and inject the knowledge to the model. We should consider fusing the prior knowledge into the diagnosis model and make the right diagnosis.
\end{itemize}

In this paper, we propose a novel multi-agent reinforcement learning (MARL) framework to solve the common thorax medical image classification problem. We design multiple prior knowledge agents to obtain the prior knowledge and the diagnostic agent to make the final diagnosis based on the actions of prior knowledge agents. 
The incremental data in the framework will also be used to update the prior knowledge learning and extraction module, which can be learned from source or target domain data and output in the same visual and embedding formats as the prior knowledge. The diagnostic agent can be seen as the main agent in our proposed framework. It will be used to handle the diagnostic problem of thoracic diseases based on prior knowledge and input medical images. 

We introduce three agents in this work. 1) Semantic agent: This agent is used to provide an initial embedding for classification and to guide the learning of parameters. 2) Visual agent: This agent is used to get the coarse region of interest information in the image. We hope that it will be able to provide visual information as a prior knowledge format. 3) Diagnostic agent: This agent is used to make the final diagnosis by combining the visual information and the semantic embedding. 
Meanwhile, the reinforcement learning framework can make the action selection process more exploratory, so when our proposed framework faces a tough case, i.e., an easily-confused case, it could make a bolder diagnostic choice and may perform better. The performance of our technique is demonstrated using the famous NIH ChestX-ray14 and CheXpert datasets. 

\subsection{Contribution}
The key contributions of our work are followed as:
\begin{itemize}
\item We propose a novel multi-agent reinforcement learning framework that can address the abnormality classification problem. This universal framework can introduce the prior knowledge to guide the diagnostic agent to improve the final performance of diagnosis;
\item We propose a uniform multi-information fusion module based on a transformer to solve the correlation among agents. We also propose a exploration-added model to handle the multi-label classification problem and high inter-class similarity problem;
\item We validated the effectiveness of our method based on some popular datasets. Several current efficient methods are used for comparison, the final experimental findings show that our method is superior;
\end{itemize}

The remainder of our paper is outlined below. In Section \ref{sec:related work}, we present the related work. The proposed solution is presented in Section \ref{sec:approach}. In Section \ref{sec:experiment}, we present key experiments. This section also includes a summary of the experimental results, and we will demonstrate the effectiveness of our approach by using it to solve a variety of thorax diseases classification problems. Implementation details are in Section \ref{sec:detils}. Finally, in Section \ref{sec:conclution}, we draw a conclusion for this work and outline future work possibilities. The supplementary materials are attached.

\section{Related Works}
\label{sec:related work}
\subsection{CXR Image Classification}
The release of some large-scale CXR datasets with more than one hundred thousand images, like NIH ChestX-ray14\citet{wang2017chestx} drives the development of data hungry deep learning for the task of CXR image analysis. Besides the CXR dataset we mentioned above, CheXpert\citet{irvin2019chexpert} is another large-scale chest radiograph dataset that is widely used in CXR image processing. 
Rocha \textit{et al.}\citet{rocha2022attention} trained a spatial transformer network on CheXpert, without the need for localization labels. 
Liu \textit{et al.}\citet{liu2019sdfn} argued that traditional deep learning approaches dealing with the presence of potentially misaligned or unrelated objects throughout the CXR image may lead to unnecessary noise, and the restriction of image resolution may lead to the loss of image details so it is difficult to detect the pathology with small lesion area, so they proposed a framework trained with high resolution images and utilizing domain knowledge at the same time.
Saleem \textit{et al.}\citet{saleem2021classification} used a transfer learning technique to detect tuberculosis and achieved a decent result. 
Zhu \textit{et al.}\citet{zhu2022pcan} argued that existing deep networks typically use the global mean pooling layer to generate features for classifiers, but the relative size, absolute size, and location of occurrence may limit classification performance. So they proposed a pixel-wise classification and attention network, which can ameliorate the above problems.
Minaee \textit{et al.}\citet{minaee2020deep} used different types of CNN-based backbones to detecte lung lesions in COVID-19 patients, while Park \textit{et al.}\citet{park2022multi} utilized transformer-based method to deal with similar problems. 
Ke \textit{et al.}\citet{ke2021chextransfer} investigated different ImageNet\citet{deng2009imagenet} pretrained architectures’ performance and parameter efficiency on CheXpert. 
Paul \textit{et al.}\citet{paul2021discriminative} proposed a few-shot CXR diagnostic method and introduced a saliency-based classifier to extract features from the output of the CNN and then classify.
In this paper, to solve the CXR image classification task, we hope to propose a model as simple as possible without degrading the classification performance.

\subsection{Multi-label Classification}
Multi-label classification is a relatively more challenging task compared with single-label classification, in which each sample may have more than one associated label. Recently, multi-label classification has attracted much research attention. 
Zhang \textit{et al.}\citet{zhang2020large} proposed a unified deep learning framework for the scenario of multi-label unknown image classification, whose results are comparable to most relevant methods. Class imbalance problem is the main challenge of multi-label classification, which means partly categories occur more frequently in the data space than others\citet{tarekegn2021review} , Jain \textit{et al.}\citet{jain2017addressing} noticed that category imbalance is common in clinic diagnosis and classifier tends to neglect the important impact of the minority category, therefore they proposed an algorithm to help biased classifier to perform well on minority class.

How to model relations between labels is another challenge in the multi-label classification task, as natural images are intrinsically and frequently relevant and co-occurrence. 
The author, in\citet{sun2017addressing} proposed a novel framework to explore the unseen relationship between labels and address the class imbalance problem at the same time. 
Wu \textit{et al.}\citet{wu2018cost} mentioned a cost-sensitive model to tackle the issue that traditional multi-label classification tasks usually ignore the label correlations by using binary relevance\citet{boutell2004learning} to find associations between labels. The cost-sensitive loss in the model deals with the label imbalance problem at the same time. 
Chen \textit{et al.}\citet{chen2019multi} considered the co-occurrence of different objects in different images and proposed a framework to model the label relationship based on Graph Convolution Network (GCN)\citet{kipf2016semi}.

\subsection{Transformer In Vision Tasks}
Transformer\citet{vaswani2017attention} was initially applied in natural language processing(NLP) tasks, Devlin \textit{et al.}\citet{devlin2018bert} proposed a model named BERT based on Transformer and obtained a prominent progress on 11 NLP tasks.
Kant \textit{et al.}\citet{kant2018practical} proposed a model to classify the multi-emotion sentiment by training a Transformer encoder-decoder architecture, Vila \textit{et al.}\citet{vila2018end} proposed an attention-based end-to-end Transformer model to translate Spanish to English and performed well. 
Recently, the huge application prospect of Transformer has been developed for computer vision tasks. Dosovitskiy \textit{et al.}\citet{dosovitskiy2020image} proposed the Vision Transformer (ViT), which is directly applied to image patches to classify the images. Liu \textit{et al.}\citet{liu2021swin} proposed a novel hierarchical vision Transformer model named Swin Transformer, which could be used as an alternative option when choosing a computer vision task's backbone. 
Carion \textit{et al.}\citet{carion2020end} proposed an end-to-end object detection model named DETR based on Transformer and achieve a prominent detection result.

\subsection{Reinforcement Learning for Medical Image Processing}
Reinforcement Learning simply means that agents choose the appropriate behavior in order to maximize the reward in a specific dynamic environment which can naturally be applied in games such as atari\citet{mnih2013playing}.
Besides, reinforcement learning for medical image processing has developed rapidly in recent years, Dou \textit{et al.}\citet{dou2019agent} proposed a RL based framework to deal with the fetal brain tissue location task, Liao \textit{et al.}\citet{liao2020iteratively} proposed a relative model with cross-entropy gain based reward to realise 3D medical image segmentation task by treating each voxel as an agent, making it a MARL problem naturally. Considered that the processing of medical images might require large number of labeled images, Stember \textit{et al.}\citet{stember2020deep} proposed a RL based method by training a deep Q network(DQN)\citet{mnih2013playing} to detect brain lesions on MRI. 
Ghesu \textit{et al.}\citet{ghesu2016artificial} proposed an agent learning model based on RL for anatomical landmark detection in medical images by considering image parsing as a policy encoding problem. 
Hou \textit{et al.}\citet{hou2021automatic} proposed a CXR report-generation framework using cascaded encoders, decoders, and a reward module. The proposed method combined adversarial training with RL and considered both results and literature fluency. 
In order to overcome the obstacles of traditional machine learning methods in dealing with anatomy detection, such as the use of computationally sub-optimal search schemes, Ghesu \textit{et al.}\citet{ghesu2017multi} proposed a RL-based framework combined with multi-scale analysis that considered the 3D detection task as a behavior learning task, training a well-designed agent to detect the abnormal area in CT images. 

\begin{figure*}
\centering
\includegraphics[width=0.95\linewidth]{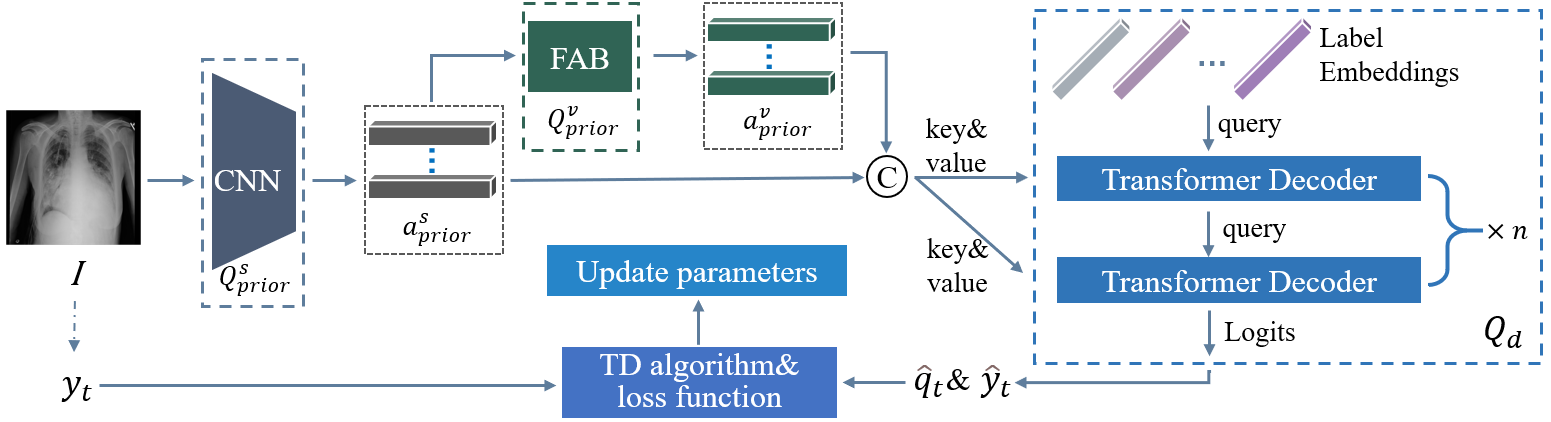}
\caption{The framework of our approach. The input is medical image $I$. This framework includes two prior knowledge agents $Q_{prior}^v$, $Q_{prior}^s$ and a diagnostic agent $Q_d$. The related actions are $a_{prior}^v$, $a_{prior}^s$ and $a_d$. The environment provides the ground truth $y_t$. Given a quadruple notation ($s_t,a_t,r_t,s_{t+1}$), the prediction value of DQN is $\hat{q}_t$, TD target is $\hat{y}_t$, then update the parameters in DQN via TD algorithm and loss function.}
\label{framework}
\end{figure*}

\section{Our Approach}
\label{sec:approach}
In this section, we first introduce the problem definition and illustrate the details of our approach in the next subsections.

\subsection{Problem Definition}
In this study, the diagnostic process can be seen as MDP then handled using the RL framework. In each stage of diagnosis, the doctor should make an accurate diagnosis based on the medical image and the doctor's prior knowledge like Fig.\ref{motivation}.
Based on this analysis, we apply the multi-agent reinforcement learning framework, which includes multiple prior agents and one main agent. The framework is shown in Fig.\ref{framework}. The prior agents are used to extract the prior knowledge and the main agent make the final diagnosis based on prior agents' actions. 
The details of definitions are followed as:
\begin{itemize}
\item \textbf{Agent}: We propose a prior knowledge agent set, $Q_{p}=\{Q_{p}^1,Q_{p}^2,...,Q_{p}^n\}$. They can provide different formats of prior knowledge to help the main agent make an accurate diagnosis. The main agent is used to judge the disease categories; $Q_d$ is defined as the diagnostic agent (main agent). 
In this work, we define two prior knowledge agents, $Q_{p}^s$ and $Q_{p}^v$, which represent semantic agent and visual agent, respectively. We will detail these agents in the next few subsections.
\item \textbf{Action}: 
According to prior agents $Q_{p}^v$ and $Q_{p}^s$, we define the $a_{p}^v$ and $a_{p}^s$ as their action respectively. For the main agent, we define $a_d$ as the final diagnostic action selection. The visual agent $Q_{p}^v$ is utilized to locate the area of disease in the medical image $I$. The action $a_{p}^v$ is the coarse semantic segmentation image. The actions $a_{p}^s$ is the preliminary classification representation, which can be obtained by pre-trained modal.
\item \textbf{Environment}: The environment $\mathcal{E}$ provides the influence based on the action of agents, the influence will also be used to update the visual agent and the semantic agent. 
The prior knowledge agents $Q_p$ utilize the input query CXR images $I$ and its associated label as its environment, the label embeddings inside $Q_d$ are initialized by the prior knowledge of $Q_p$. Moreover, when we get the TD target, it also interacts with the prior agents and diagnostic agents, so the environment is dynamically interactive in our setting.

\item \textbf{State}: The state is the feature vector $s$ of input CXR image in each stage. Here, we apply the classic CNN model to extract $s$, which is trained by old data or other domain's similar data. 
\item \textbf{Reward}: 
We define the $r^d$ as the reward of main agent $Q_d$ and $r_i^p$ as the reward of prior agent $Q_p^i$. Discount return is discount accumulative reward. The discount return of main agent is $U_t$=$\sum_{j=0}^{\infty} {\gamma}^j\cdot r_{t+j}^d$, where {$\gamma{\in[0,1)}$} is the discount factor, $t$ is the index of training episode. The discount return of prior agent is $U^{p_i}_t$=$\sum_{j=0}^{\infty} {\gamma}^j\cdot r_{t+j}^{p_i}$,
For $Q_d$ and $Q_p$, we define that when a specific pathology is correctly classified, $r^d$ and $r^q_i$ equal to $1$ else $-1$.
\end{itemize}

Based on these definitions, our goal is to seek a target $Q^*_d(s_t,a_t^v,a_t^s)$ function, which can provide the final accurate diagnosis, where $a_t^v$ and $a_t^s$ are the action of prior agents in time $t$. We also need to optimize the agent $Q^*_v(s_t)$ and $Q^*_s(s_t)$. 
Thus, the goal of RL is to maximize the expected cumulative reward:
\begin{equation}
    max_{w_s,w_v,w_d}\mathbb{E}[U_t|s_t,a_t^v,a_t^s]+\sum_{i=0}^{n}\mathbb{E}[U^{p_i}_t|s_t],
    \label{objectfunction}
\end{equation}
where $w_d$ denotes parameters of $Q_d$ function, $w_s$ and $w_v$ denote the parameters of $Q_p$, $t$ denotes the index of learning episode, $U$ denotes the expectation of reward. $n$ denotes the number of prior agents. In this paper, $n=2$. 

In our framework, we extract the feature as the state $s_t$ when we input the medical image $I$. Then, the prior agents $Q_{p}$ can select a set of action $a^v$ and $a^s$. The main agent $Q_d$ outputs the action $a_t$ based on $s_t$, $a^v$ and $a^s$. Here, we use vanilla temporal difference(TD) algorithm\citet{watkins1989learning} to train the DQN and define the loss $\mathcal{L}_{TD}$ for the RL part based on the one-step TD error.

Based on feedback of the environment, we can expect achieving the related true label. These label can also be used to fine-tune the prior agents via the classic cross-entropy loss. The optimization function \ref{objectfunction} can be rewritten as:
\begin{equation}
\begin{aligned}
    max_{w_s,w_v,w_d}&\mathbb{E}[U_t|s_t,a_t^v,a_t^s]+\sum_{i=0}^{n}\mathbb{E}[U^{p_i}_t|s_t]+\\
    &\mathbb{E}[y_t|s_t,a_t^v,a_t^s]+\sum_{i=0}^{n}\mathbb{E}[y_t|s_t],
    \label{objectfunction2}
\end{aligned}
\end{equation}
where $y_t$ is the true label of input image $I$. Here, $\mathbb{E}[U_t|s_t,a_t^v,a_t^s]$ and $\mathbb{E}[U^{p_i}_t|s_t]$ can apply the TD loss to optimization the model parameters. $\mathbb{E}[y_t|s_t,a_t^v,a_t^s]$ and $\mathbb{E}[y_t|s_t]$ can apply the classic cross-entropy loss to solve the optimization problem. Thus, the final object can be written as:
\begin{equation}
\begin{aligned}
\mathcal{L}=&\sum_{i=0}^{n}\mathcal{L}_{p}+\sum_{i=0}^{n}\mathcal{L}^p_{TD}+\mathcal{L}_{TD}+\mathcal{L}_d,
\label{obj}
\end{aligned}
\end{equation}
where $\mathcal{L}_{p}$ and $\mathcal{L}_d$ denotes the cross-entropy losses of prior agents and main agent, $\mathcal{L}^p_{TD}$ denotes the TD loss of prior agent and $\mathcal{L}_{TD}$ denotes the TD loss of main agents. $n$ denotes the number of prior agents. We will details these losses in the later implementation stage.



Until now, we design the process of model learning based on the RL framework. However, in each diagnostic case, the model only selects the action with the highest value as the final diagnosis. This condition loses the human ability to explore in some tough cases. To handle this problem and make the diagnostic model smarter and more daring, we applied the $\epsilon$-greedy\citet{watkins1989learning} algorithm to give our model the ability to explore. The key idea of this policy is with the probability $\epsilon$ to explore while with the probability (1-$\epsilon$) to use the learned knowledge, which ensures that non-optimal cases can be chosen. Well in this setting, we want to explore the non-optimal actions as our choice and make full use of our learned prior knowledge.
In order to deal with the problem of the exploration-exploitation dilemma, we choose $\epsilon$-greedy policy as our behavior policy to control the selection of actions:
\begin{itemize}
    \item $1-\epsilon$: When the probability is $1-\epsilon$, we chose the action with the highest score as the diagnostic result. Note that our target problem is a multi-label classification problem, so on the premise that the classification threshold is satisfied, each agent may choose multiple actions. 
    \item $\epsilon$: When the probability is $\epsilon$, we expect the agent can explore the different diagnosis like human. However, we do not allow this agent to randomly select the actions as the diagnosis results, which also should follow some prior knowledge. Thus, we delete the action with the highest score, and then select the remainder of the highest scores from the remaining actions as diagnostic results. 
\end{itemize}

Apparently in the whole training process, $\epsilon$ is decreasing gradually, exploration at first and exploitation after learning some useful information. In the training stage, we choose $\epsilon$ = max\{$\epsilon_{min},1-\frac{1-\epsilon_{min}*step}{total}$\}. $\epsilon$ equals to 0 when testing the model.
We will detail some key elements in our proposed framework in the next few subsections.


\subsection{Prior Agent}
\subsubsection{Semantic Agent}
The semantic agent ($Q_{p}^s$) is used to extract and provide prior knowledge as the semantic format from the query CXR image $I$. We design the semantic information extraction network based on the ResNet structure\citet{he2016deep}.
Here, we construct the 2D network with ResNet-50 and 3 additional convolutional layers with different kernel sizes, which can effectively extract the detail information from different scale of medical image. 
The action of this agent can be seen as the preliminary delineation of some pathologies.
The initialized parameter of this agent can be learned by the training data, which can also be learned by other cross-domain data or other large-scale data as the prior knowledge (experience). We will discuss its different performance in experiment section.  

By the semantic agent, we can obtain the output features $f_s^i$ $\in$ ${\mathbb{R}}^{d\times C}$, which represents the extracted feature and mainly contains information about $i^{th}$ category from medical image. We employ the classifier to process each $f_s^i$ into probability value $p_s^i$, which represents the classification result for $f_s^i$. We take $p_s^i$ as the final confidence of $i^{th}$ category. The output is composed of all categories confidences, which is denoted as $p_t$ $\in$ $\mathbb{R}^{C}$. The loss function $\mathcal{L}_{p}$ can be written:
\begin{equation}
\begin{split}
{\mathcal{L}_{p}}& = -\sum_{i=1}^{C} t_j^i\log(p_t^i),
\label{semantic}
\end{split}
\end{equation}
where $\textbf{T}_j=\{t_j^1, t_j^2, ...., t_j^C\}$ represents the label of medical image $I_j$, $t_j^i$ is a binary parameter that if the current CXR image $I_j$ belongs to category $i$, $t_j^i$ = 1, otherwise $t_j^i$ = 0.
In the test phase, we compare each value in $p_t$, and take the category with highest value as the classification result. 

\subsubsection{Visual Agent}
\begin{figure}
\centering
\includegraphics[width=0.95\linewidth]{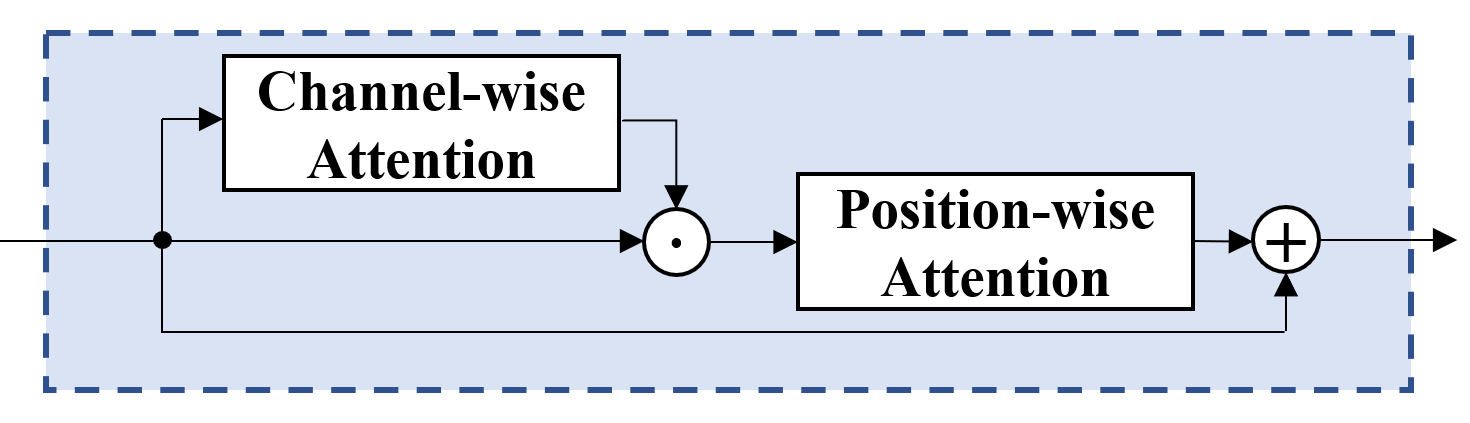}
\caption{Foreground Attention Block (FAB). We use the cascaded structure of channel- and position-wise attention to produce the foreground attention.}
\label{forground}
\end{figure}

This agent is used to extract visual attention area of the query medical image $I$ as the visual prior knowledge. The action of $Q_{p}^v$ is the foreground image feature $a_{p}^v$. The structure of this agent is shown in Fig.\ref{forground}, visual agent ($Q_{p}^v$) is used to provide prior knowledge as visual format.
This structure is inspired by the previous work \citet{ouyang2020learning}. The foreground attention block(FAB) applies a self-attention mechanism to learn and emphasize the coarse foreground attention maps. 
The channel-wise attention module is adapted from squeeze-and-excitation (SE) block\citet{hu2018squeeze} which can focus on helpful channels, and then the position-wise attention module can use channel-weighted vectors to give a preliminary and sketchy estimation of the areas.
Note that the visual agent is also trained via the loss function \eqref{semantic} and the action of the visual agent is the coarse ROI of an input CXR image.

\subsection{Diagnostic Agent}
\begin{figure}
\centering
\includegraphics[width=0.95\linewidth]{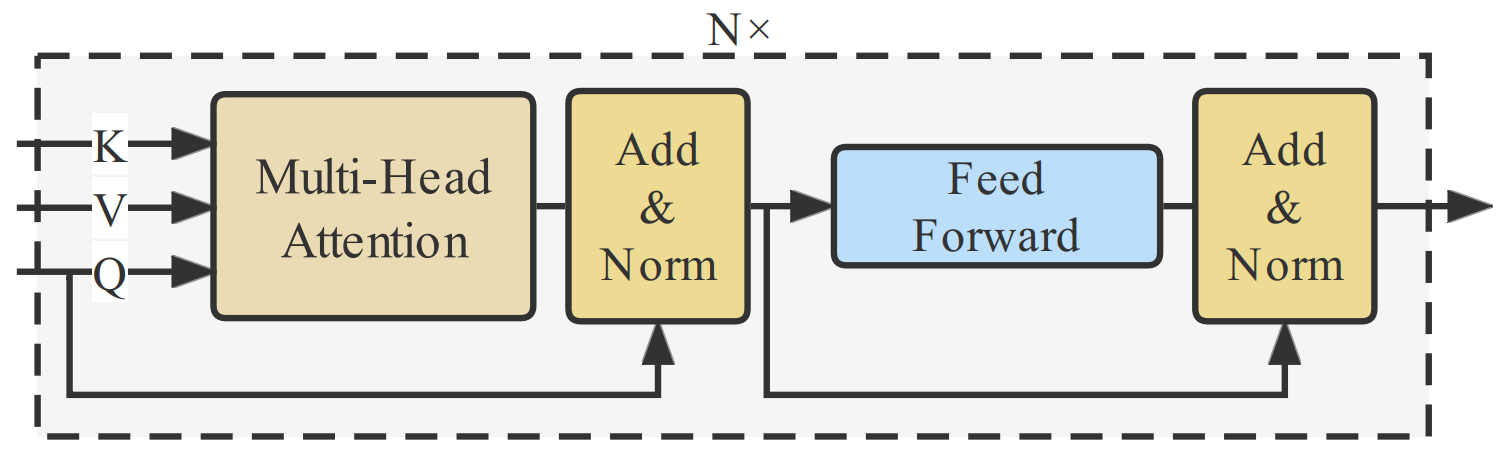}
\caption{The structure of the diagnostic agent.}
\label{decoder}
\end{figure}
This agent can be seen as the final classification agent. The classic transformer structure is applied in this agent. The goal of transformer is used to fuse the output feature of $Q_{p}^s$ and $Q_{p}^v$. Then, update the fuse feature based on the inner label embedding\citet{liu2021query2label} to achieve the final image feature representation. The detail of architecture is shown in Fig.\ref{decoder}, which is similar to the standard Transformer decoder structure without self-attention module. 
Each Transformer decoder layer $l$ updates the fused feature $F_{l-1}$ from the output of its previous layers based on the label embedding $L_l$ and spacial feature as follows:
\begin{equation}
\begin{aligned}
&q_l=L_l+p_{L_l}, k_l=f+p_{f}, v_l=f,\\
&\dot{F}_l=MultiHead(q_l,k_l,v_l)+L_{l-1},\\
&F_l=FFN(LN(\dot{F}_l))+\dot{F}_l,
\end{aligned}
\label{newv}
\end{equation}
where $l$ is the index of transformer layer, $f$ is the image feature from prior knowledge agent, $p\in \mathbb{R}^{C\times d}$ denotes the learnable position encoding. In the first layer, the input label embedding is the initialized $Q_0\in \mathbb{R}^{C\times d}$, where $C$ is number of categories and $d$ is the dimension of decoder. The output is the fused feature $F_l\in$ ${\mathbb{R}}^{C\times d}$, which is actually $a_d$ and represents the extracted feature for $C$ categories at the last layer.

Transformer decoder have a built-in cross-attention mechanism, we find common multi-label classification losses work well in dealing with CXR images classification tasks. But as we mentioned above about the data distribution imbalance problem, we want to deal with the label imbalance problem further via choosing a proper loss, then we choose an efficient version of focal loss\citet{liu2021query2label}, where we can choose different $\gamma$ values for positive and negative values. Given an input CXR images $I$, we can predict its category probabilities $p = [p_1,\ldots, p_C]^T \in {\mathbb{R}}^{C}$ using our framework. We leverage the following asymmetric focal loss\citet{ridnik2021asymmetric} to implement $\mathcal{L}_{diagnostic}(y_t|f_{cls}(a_{d}))$, the second part need to minimize in the objective function \eqref{obj} to calculate the loss for each training CXR image $I$:
\begin{equation}
\mathcal{L}_{d}=\frac{1}{C}\sum^{C}_{c=1}
    \begin{cases}
    {(1 - p_c)}^{\gamma+}\log(p_c), & y_c = 1,\\
    {(p_c)}^{\gamma-}\log(1- p_c), & y_c = 0,
    \end{cases}
\label{aslloss}
\end{equation}
where $y_c$ equals to 0 when the processed CXR image has label $c$ and equals to 1 otherwise. $\gamma$+ is set to 0 and $\gamma$- is set to 1 in our experiment. When completely training all the CXR images in our chosen training data set, we average \eqref{aslloss} to calculate the loss to optimize the module. Furthermore, as we define the $Q_d$ as the main agent and the DQN in our framework, we also optimize its parameters via the TD algorithm:
\begin{equation}
\mathcal{L}_{TD}=\frac{1}{2}[Q_d(s_t,a_s, a_v, \boldsymbol{w_t^d})-\hat{y}_t]^2,
\label{td}
\end{equation}
\begin{equation}
\mathcal{L}^p_{TD}=\frac{1}{2}[Q_p(s_t,\boldsymbol{w^p_t})-\hat{y}^p_t]^2,
\label{tdp}
\end{equation}
where, $\boldsymbol{w_d}$ and $\boldsymbol{w_p}$ are the network parameter of main agent and prior agent. $\hat{y_t}=r_t+\gamma\cdot\mathop{max}\limits Q_d(s_{t+1},a;\boldsymbol{w_{t-1}})$ is the TD target. $\hat{y}^p_t$ has the same equation. At this point, the details of each part of the total loss function \eqref{obj} have been described.
Similar to the work in \citet{mnih2013playing}, we store the experiences $e_t=(s_t,a_t,r_t,s_{t+1})\in \mathcal{D}$ that agents learned at each time-step by experience reply, then sample the experiences from reply buffer by Q-learning updates during the inner loop of, after that agents can choose an action based on the $\epsilon$-greedy policy.
The full algorithm is presented in Algorithm \ref{algorithm23}.

\begin{algorithm}[ht]
\caption{MARL with DQN}
\begin{algorithmic}[1]
\Require Image data $I$ and corresponding ground truth $y_t$, initialize learnable parameters $\boldsymbol{w_s}$ of semantic agent, $\boldsymbol{w_v}$ of visual agent, $\boldsymbol{w_d}$ of diagnostic agent with random weights
\Ensure Initialize reply memory $\mathcal{D}$ to capacity $\mathcal{N}$

\State Prior knowledge agents select actions and update parameters $\boldsymbol{w_s}$ and $\boldsymbol{w_v}$ by the image data $I$ and its related ground truth $y_t$ based on the loss function \eqref{semantic} and \eqref{tdp}
\For{episode=1, $M$}
    \State Initialise $s$
\For{$t=1, T$}
    \State Diagnostic agent select an action $a_d$ via $\epsilon$-greedy
    \State Execute action $a_d$ and observe reward $r_d$
    \State Set $s_{{t+1}\_d}$=$s_{t\_d}$
    \State Store ($s_{{t}\_{d}}$,$a_{{t}\_{d}}$,$r_{{t}\_{d}}$,$s_{{t+1}\_{d}}$) as a transition in $\mathcal{D}$
    \State Sample random minibatch of transitions 
    \Statex \qquad\qquad ($s_{{j}\_{d}}$,$a_{{j}\_{d}}$,$r_{{j}\_{d}}$,$s_{{j+1}\_{d}}$) from $\mathcal{D}$
    \State Get $\hat{y}_{j\_d}$ base on TD target
    \State Update $\boldsymbol{w_d}$ by \eqref{td} and \eqref{aslloss}
    \State Fine-tune $\boldsymbol{w_s}$ and $\boldsymbol{w_v}$
\EndFor
\Statex \qquad \textbf{end for}
\EndFor
\Statex \textbf{end for}
\Statex \textbf{Return:} Updated DQN parameters $\boldsymbol{w_{\{s,v,q\}}}$ via \eqref{obj}
\end{algorithmic}
\label{algorithm23}
\end{algorithm}

\section{Experiment}
\label{sec:experiment}
\subsection{Data Sets}
We evaluate our method on the publicly released NIH ChestX-ray14 and CheXpert datasets. NIH ChestX-ray14 consists of 112,120 frontal-view X-ray images of 30,805 unique patients, and we report the 14 thorax abnormal diseases classification performance on the testing set. 
CheXpert is another large-scale dataset for chest X-rays released by Stanford University. It contains 224,316 chest radio graphs of 65,240 unique patients. 14 observations are labeled in radiology reports, capturing uncertainties inherent in radiography interpretation. Part of the CXR images is a frontal view, and the rest of the data set is a lateral view. 

\subsection{Evaluation Metrics}
In our study, we chose the area under the receiver operating characteristic (ROC) curve, namely the AUC to evaluate the performance of the abnormal thorax disease classification. We conduct some experiments to test our proposed model's performance compared with the previous state-of-the-art (SOTA) framework. The AUC score of each pathology and the average AUC score over all pathologies are reported, respectively.

\subsection{Implementation Details}
\label{sec:detils}
We use PyTorch for implementation. We adopt vanilla Resnet-50\citet{he2016deep} as our backbone. 
Our experiment is operated by using NVIDIA GeForce RTX 3090 with 24GB memory. For training, we perform data augmentation by using RandAugment\citet{cubuk2020randaugment} and normalizing CXR images with zero mean and one std rather than mostly used ImageNet\citet{deng2009imagenet} pretrained images' mean and std according to the nature of CXR images. We use the Adam\citet{kingma2014adam} optimizer with True-weight-decay\citet{loshchilov2017decoupled}1$e$-2 and maximal learning rate is 1$e$-3, and we choose initial $\epsilon_{min}$ equals to 0.2 during the training step. At the same time, we resize the original images to and 512 $\times$ 512 and randomly horizontal flipping.

\subsection{Comparison with State-of-the-art Methods}

\subsubsection{NIH ChestX-ray14}
We first compared with some SOTA methods on the NIH ChestX-ray14 dataset.
The AUC score of each pathology is summarized in Table.\ref{tb1}. The average AUC score of our baseline arrives at 0.866 across the 14 thoracic diseases. It is competitive with or better than the previous works. In this experiment, we do not apply additional data to train the prior agents in order to make sure the comparison experiment is fair. 
\begin{table*}[t]
\setlength\tabcolsep{3pt}
  \centering
  \caption{Comparation Of AUC Scores with previous SOTA Works on the official split of NIH ChestX-ray14. We report the AUC with 95 \% confidence interval (CI) of our method.The best performance of each pathology is shown in bold.}
    \begin{tabular}{c|c|c|c|c|c}
    \toprule
    \tabincell{c}{Abnormality}&\tabincell{c}{Wang \textit{et al.}\\\citet{wang2017chestx}  }&\tabincell{c}{Xi \textit{et al.}\\\citet{ouyang2020learning}}&\tabincell{c}{ImageGCN\\\citet{mao2022imagegcn}}&\tabincell{c}{DGFN\\\citet{gong2021deformable}}&\tabincell{c}{Ours}\\
    \midrule
    Atelectasis & 0.70  & 0.77  & 0.80 & 0.82 & \textbf{0.83} (0.82, 0.84) \\
    \midrule
    Cardiomegaly & 0.81   & 0.87  & 0.89 &\textbf{0.93} & \textbf{0.93} (0.91, 0.94) \\
    \midrule
    Effusion & 0.76   & 0.83  & 0.87 &0.88 & \textbf{0.90} (0.89, 0.92) \\
    \midrule
    Infiltration & 0.66   & 0.71  & 0.70 & \textbf{0.75} & \textbf{0.75} (0.74, 0.76) \\
    \midrule
    Mass  & 0.69    & 0.83  & 0.84 & 0.88 & \textbf{0.89} (0.88, 0.91) \\
    \midrule
    Nodule & 0.67   & 0.79  & 0.77 & 0.79 & \textbf{0.81} (0.80, 0.82)  \\
    \midrule
    Pneumonia & 0.66    & \textbf{0.82}  & 0.72 & 0.78& \textbf{0.82} (0.81, 0.83)  \\
    \midrule
    Pneumothorax & 0.80     & 0.88  &0.90  &0.89 & \textbf{0.91} (0.90, 0.93) \\
    \midrule
    Consolidation & 0.70    & 0.74  & 0.80 & 0.81 & \textbf{0.82} (0.81, 0.85) \\
    \midrule
    Edema & 0.81    & 0.84  & 0.88 &0.89 & \textbf{0.91} (0.90, 0.92) \\
    \midrule
    Emphysema & 0.83    & \textbf{0.94}  & 0.92 & \textbf{0.94} & \textbf{0.94} (0.94, 0.95)  \\
    \midrule
    Fibrosis & 0.79    & 0.83  & 0.83 & 0.82 & \textbf{0.85} (0.84, 0.86)  \\
    \midrule
    Pleural\_Thickening & 0.68   & 0.79  & 0.79 & 0.81 & \textbf{0.83} (0.81, 0.84) \\
    \midrule
    Hernia & 0.87  & 0.91  & \textbf{0.94} & 0.92 & \textbf{0.94} (0.93, 0.95) \\
    \midrule
    Mean AUC & 0.745  & 0.819  & 0.832 &0.850 & \textbf{0.866} (0.859, 0.873)\\
    \bottomrule
    \end{tabular}%
  \label{tb1}%
\end{table*}%

From these results, we have the following observations:
\begin{itemize}
    \item The results obtained by our method are consistent with the optimal algorithms for ``Cardiomegaly'', ``Infiltration'', ``Emphysema'', ``Pneumonia'' and ``Hernia''. The last two pathologies are the least frequent of all pathologies, they account for about only 1.1\% percent of the total data. All of the compared methods perform poorly in these two pathologies, which can be attributed to the long tail distribution problem. In order to solve above problem, our method guides the multi-information for the final pathology representation with the help of prior agents. The goal of our approach is the pursuit of optimal overall agents. It demonstrates the effectiveness of multi-agent framework indirectly. 
    \item The four diseases, ``Cardiomegaly'', ``Emphysema'', ``Edema'', and ``Fibrosis'' account for only 6.6\% of total pathologies, but we achieve SOTA AUC compared with other previous work. 
    The reason is that we utilize the exploratory part of our MARL framework to give our model the opportunity to try some more reasonable choices, then solve the long tail distribution problem effectively.
    Meanwhile, the model of prior knowledge can provide effective auxiliary information and additional experience. Other comparable methods mostly rely on the quality of training data, but when they train with fewer samples, they are often powerless. 
\end{itemize}

\subsubsection{CheXpert}
We also report the performance of our proposed MARL framework on the CheXpert data set in this section. We need to note that the data structure in CheXpert is different from NIH ChestX-ray14 data set, so we based on the baseline work\citet{irvin2019chexpert}, map the uncertain labels to 1 as it works well and make the proposed model outputs the maximum probability of the observations across the views. Meanwhile, based on the baseline, we also choose to evaluate 5 classes which called competition tasks, which selected according to clinical importance and prevalence. Our results are shown in Table.\ref{X} and we adopt different backbones. Same as the experiment on NIH ChestX-ray14, when Resnet is used as the backbone we got the better results. We present the performance obtained by a single model rather than the performance of Irvin \textit{et al.}\citet{irvin2019chexpert}, which is from the ensemble of 30 models. In our proposed framework, the average AUC scores over five pathologies achieve 0.925. 

In this data set, when using our framework and the uncertain labels are set to “ones”, we get the mean AUC of 0.925, while when using our framework and the uncertain labels are set to “zeros”, the AUC of each pathology is 0.831, 0.877, 0.933, 0.902, 0.938, respectively, the average AUC score is 0.8962, which surpasses the corresponding baseline ``U-Zeros''. Here, we notice that our model shows its superiority for some pathologies except ``Edema'', the performance of ``Cardiomegaly'' and ``Atelectasis'' are significantly improved (0.832 vs 0.890, 0.858 vs 0.922), which improved 6.97\% and 7.46\%, respectively. 
\begin{table*}[t]
\renewcommand\arraystretch{0.3}
\setlength\tabcolsep{4pt}
  \centering
  \caption{Comparison of AUC scores with different models on CheXpert validation set. Our approach shows the results under different backbone, DenseNet and ResNet.}
    \begin{tabular}{c|c|c|c|c|c|c}
    \toprule
    \tabincell{c}{Abnormality}&\tabincell{c}{U-Zeros\\\citet{irvin2019chexpert}}&\tabincell{c}{U-Ones\\\citet{irvin2019chexpert}}&\tabincell{c}{CT+LSR\\\citet{pham2021interpreting}}&\tabincell{c}{Xi \textit{et al.}\\\citet{ouyang2020learning}}&\tabincell{c}{Ours\\(DenseNet)}&\tabincell{c}{Ours\\(ResNet)}\\
    \midrule
    Atelectasis   & 0.811  & 0.858  & 0.825 &0.920 &0.916 & \textbf{0.922} \\
    \midrule
    Cardiomegaly   & 0.840  & 0.832  & 0.855 &0.886 &0.879 & \textbf{0.890} \\
    \midrule
    Consolidation  & 0.932  & 0.899  & 0.937 &0.907 &0.936 & \textbf{0.942}  \\
    \midrule
    Edema   & 0.929  & \textbf{0.941} & 0.930 &0.937& 0.934 & 0.932  \\
    \midrule
    Pleural Effusion   & 0.931  & 0.934  & 0.923 &0.933 &0.935 & \textbf{0.938} \\
    \midrule
    Mean AUC  & 0.889 & 0.893 & 0.894 &0.917 &0.920 & \textbf{0.925} \\
    \bottomrule
    \end{tabular}%
  \label{X}%
\end{table*}%

\subsection{Few-shot and Data-efficiency Problem}
We conduct an experiment on the NIH ChestX-ray14 data set to demonstrate that our proposed MARL framework can handle the data-efficient and few-shot learning problem. 

\begin{figure}[t]
\centering
\includegraphics[width=0.95\linewidth]{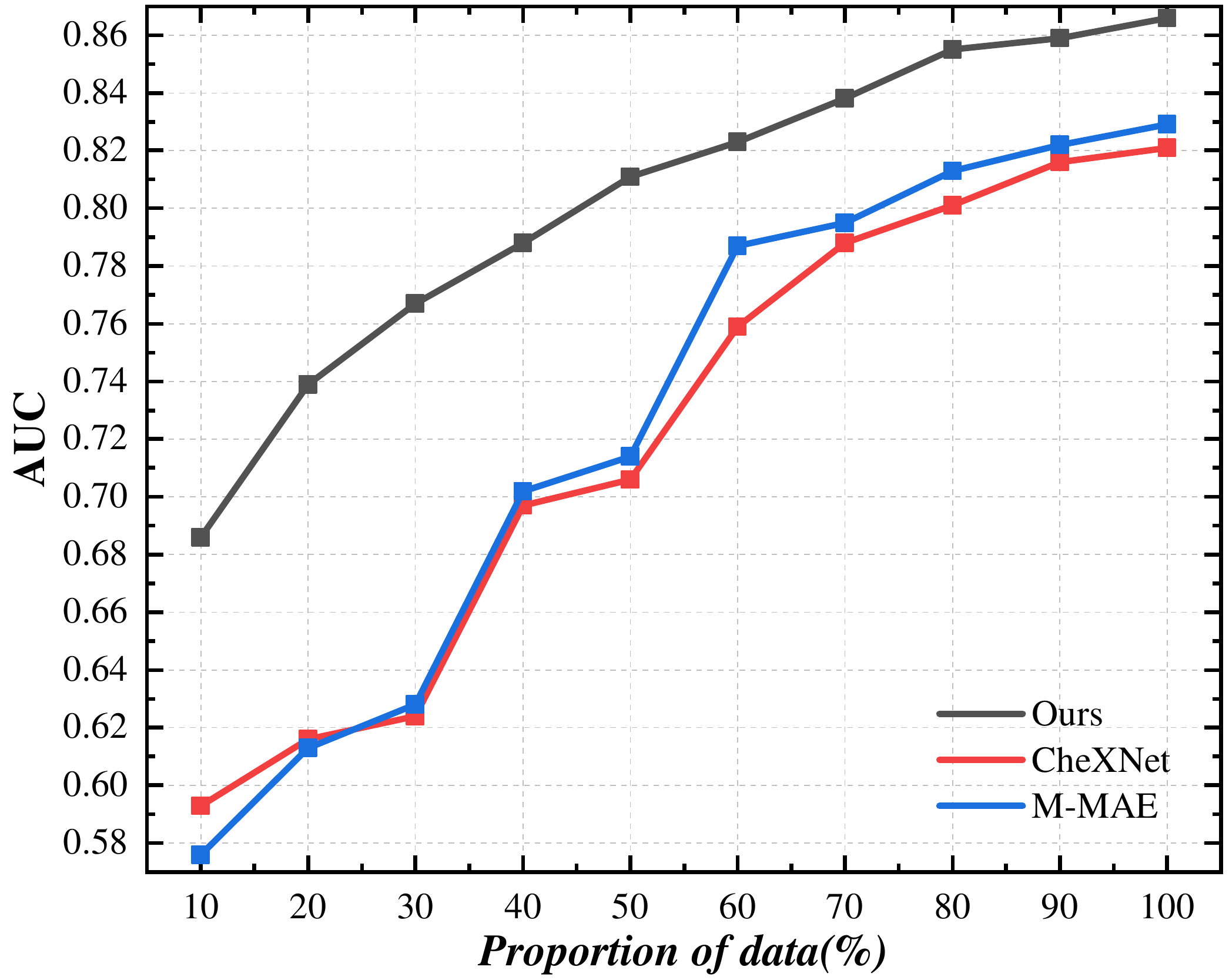}
\caption{The process of prior knowledge accumulation and illustration of data-efficient. The ordinate is AUC scores while the abscissa means the percentage of total data volume of CXR images a doctor has learned. We denote CheXNet and M-MAE as the method in \citet{seyyed2020chexclusion,xiao2023delving} respectively.} 
\label{few-shot}
\end{figure}
In terms of the few-shot learning problem, we split the NIH dataset, in which 9 types of pathologies with about 1,100 samples are chosen as auxiliary set and the remaining 5 types of samples are taken as the test set. We perform 5-way 1-shot, 5-way 5-shot, and 5-way 10-shot classification tasks on the NIH data set. Take 5-way 5-shot as an example: there are 5 different pathology categories with 5 support images and 10 query images per category, \textit{i.e.} 5$\times$5 + 10$\times$5 = 75 images in each episode. 
We pre-trained the prior agent using the CheXpert dataset. The two datasets do not have the same category. As a result, we can only guarantee dimensionality consistency in the initial guided learning. Naturally, we then apply fine-tuning at the next time. To divide the query images into the closest category, we use a Euclidean and Cosine distance metric, as well as a softmax function.

Compared with other methods using few-shot learning \citet{jin2021ctfc,jiang2022multi,finn2017model,gordon2018versa} as shown in Table.\ref{nwks}, we achieve the acc of 44.26, 49.63, and 52.64 in the 1, 5, and 10-shot scenarios. Experimental results prove that the prior knowledge learning module and the reinforcement learning setting in this framework can help to solve the few-shot learning problem, and the exploration ability of our method can fit the nature of humans, which helps to improve model performance.
At the same time, when we train the model with the same scale of data as other few-shot learning methods, we can also achieve better classification performance. 
\begin{table}[t]
  \centering
  \caption{The 3-way, 1, 5, 10-shot classification accuracy(\%). The best results are highlighted.}
    \begin{tabular}{c|c|c|c}
    \hline
    FSL Methods & 1-shot & 5-shot & 10-shot \\
    \hline
    MAML\citet{finn2017model}& 39.96 & 46.83   & 48.69 \\
    Versa\citet{gordon2018versa}& 37.94 & 45.81   & 47.13 \\
    M-Learner\citet{jiang2022multi} & 43.37   & 47.95 &50.58 \\
    \hline
    Ours& \textbf{44.26} & \textbf{49.63}   & \textbf{52.64} \\
    \hline
    \end{tabular}%
  \label{nwks}%
\end{table}%

We also modify the number of training data to further observe the impact of the number of training data on model performance. The related experimental results are shown in Fig.\ref{few-shot}, at the same time, we reproduce two other works \citet{xiao2023delving,seyyed2020chexclusion} to compare with our proposed method on the matter of data efficient. From these results, we can find that we train the diagnostic agent with only about 80\% of CXR images, the mean AUC of the framework is comparative with previous SOTA work. Besides, we can find that when selecting the same amount of data for training, our results always outperform than the other two methods. We think the reason is that the RL framework allows each training to have more exploration capabilities, allowing the model to perform reasoning like a person, thereby amplifying the role of training data. 

\subsection{Ablation Study}
In this section, we conduct the ablation study to prove the superiority of our proposed MARL thorax diseases classification framework. We run the entire ablation experiment on the NIH ChestX-ray14 data set to demonstrate the significance of each agent and the MARL setting. 
The related experiments are shown in Table.\ref{abl}. Note that only model 8 remove the RL-related settings. ``+'' represents the framework using that element, while ``-'' means the framework removing the element. Model 9 which contains three ``+'' is our proposed integrated MARL framework, while Model 8 is a model based on Model 9 but removes all the MARL-related settings. We can conclude that the prior knowledge extraction module of our proposed framework is important to help the diagnostic agent make a decision. We can also confirm the diagnostic agent is good at decoupling and fusing prior knowledge from the visual and semantic agents. Furthermore, the RL-related setting is key to making a more accurate diagnosis.
\begin{table}[htbp]
\setlength\tabcolsep{4pt}
  \centering
  \caption{Ablation study. Note that model 8 removes the reinforcement learning settings, compared with model 9.}
    \begin{tabular}{c|c|c|c|cr}
\cmidrule{1-5}
\tabincell{c}{Model}&\tabincell{c}{Visual\\Agent}&\tabincell{c}{Semantic\\Agent}&\tabincell{c}{Diagnostic\\Agent}&\tabincell{c}{AUC}\\
\cmidrule{1-5}   
    1     & -     & +     & +     & 0.862     &  \\
    2     & -     & -     & +     & 0.856     &  \\
    3     & -     & -     & -     & 0.839     &  \\
    4     & -     & +     & -     & 0.850     &  \\
    5     & +     & +     & -     & 0.838     &  \\
    6     & +     & -     & +     & 0.861     &  \\
    7     & +     & -     & -     & 0.841     &  \\
    8     & +     & +     & +     & 0.837     &  \\
    9     & +    & +     & +     & \textbf{0.866}     &  \\
\cmidrule{1-5}    \end{tabular}%
  \label{abl}%
\end{table}%

Based on this setting, we can achieve the following observations:
\begin{itemize}
    \item The visual agent is the only used part in model 7, and we can find the result is just a little higher than the worst in model 3, which is easy to understand. If none of the components are used, the whole structure is totally not integrated into our proposed model 9. In the case of model 7, we only focus on the foreground information of the input CXR images and ignore the detailed information under the CXR images, thus the results are not good. But we can find that the result is still better than part of the previous work, so the result still demonstrates the performance of the visual agent.
    \item The semantic agent brings some performance improvements according to model 5, 7 and 1, 2. The results show that a strong spatial feature extractor is important and can get detailed information about the features of CXR images. According to the results of models 4, 5, and 7, we can find that when we use both the visual agent and the semantic agent, the model's performance is worse than the model using only the semantic agent. It means the foreground attention block introduces some redundant information and then impairs the final classification results. Furthermore, we can see that the model with only a semantic agent outperforms some previous SOTA works because it validated the semantic agent's validity.
    \item According to the results of models 6 and 7, the diagnostic (main) agent takes advantage of the transformer decoder structure and achieves a significant improvement over the visual and semantic modules. 
    Compared with the other models, model 9 achieves a new state-of-the-art AUC of 0.866, which represents a 2\% improvement and is better than almost all of the previous works. In this case, the prior agent, trained with all training data, is crucial to the final diagnostic result. That seems a little unfair. In order to reduce this interference, we randomly initialized the prior agent and updated it with the new data. It also achieves the mean AUC of 0.856 and also better than all of the other baseline models. 
    \item The proposed MARL framework is superior to the traditional deep learning framework in Model 8. 
    In order to demonstrate the performance of the MARL framework, we introduce Model 8, which is the model in which we remove the multi-agent reinforcement learning related settings such as action, the temporal difference algorithm, and so on. It means that we only applied the $L_p$ and $L_d$ losses to optimize the model's parameters. It achieves an AUC of 0.837, which is about 3\% lower than the integrated proposed model 9. This result demonstrates the effectiveness of the RL mechanism in our approach; when we remove it, the model 8 cannot benefit from the interaction and exploration mechanisms in reinforcement learning. Besides, the results in model 8 is still competitively compared to previous methods, it also proves the rationality of our proposed model structure.
\end{itemize}

In general, our proposed framework includes all modules, which gives us the best mean AUC of 0.866. The corresponding experiments and analyses demonstrate the performance of these modules.

\subsection{Prior Knowledge from Other Domain}
In this paper, we introduce the prior knowledge information to guide the final diagnosis. The prior agent can be seen as the Large-scale pre-trained models (PTMs). If this prior knowledge comes from data in other domains, will it have new effects on the main agent? This question is somewhat similar. If we find an orthopedic doctor to see a chest disease, will the orthopedic experience play a positive role?
In order to explore this problem, we applied the other data set to initialize the parameters of the visual agent and the semantic agent. Then, we trained the main agent and fine-tuned the prior agents based on the NIH ChestX-ray14 data set.

\begin{figure}
\centering
\includegraphics[width=0.95\linewidth]{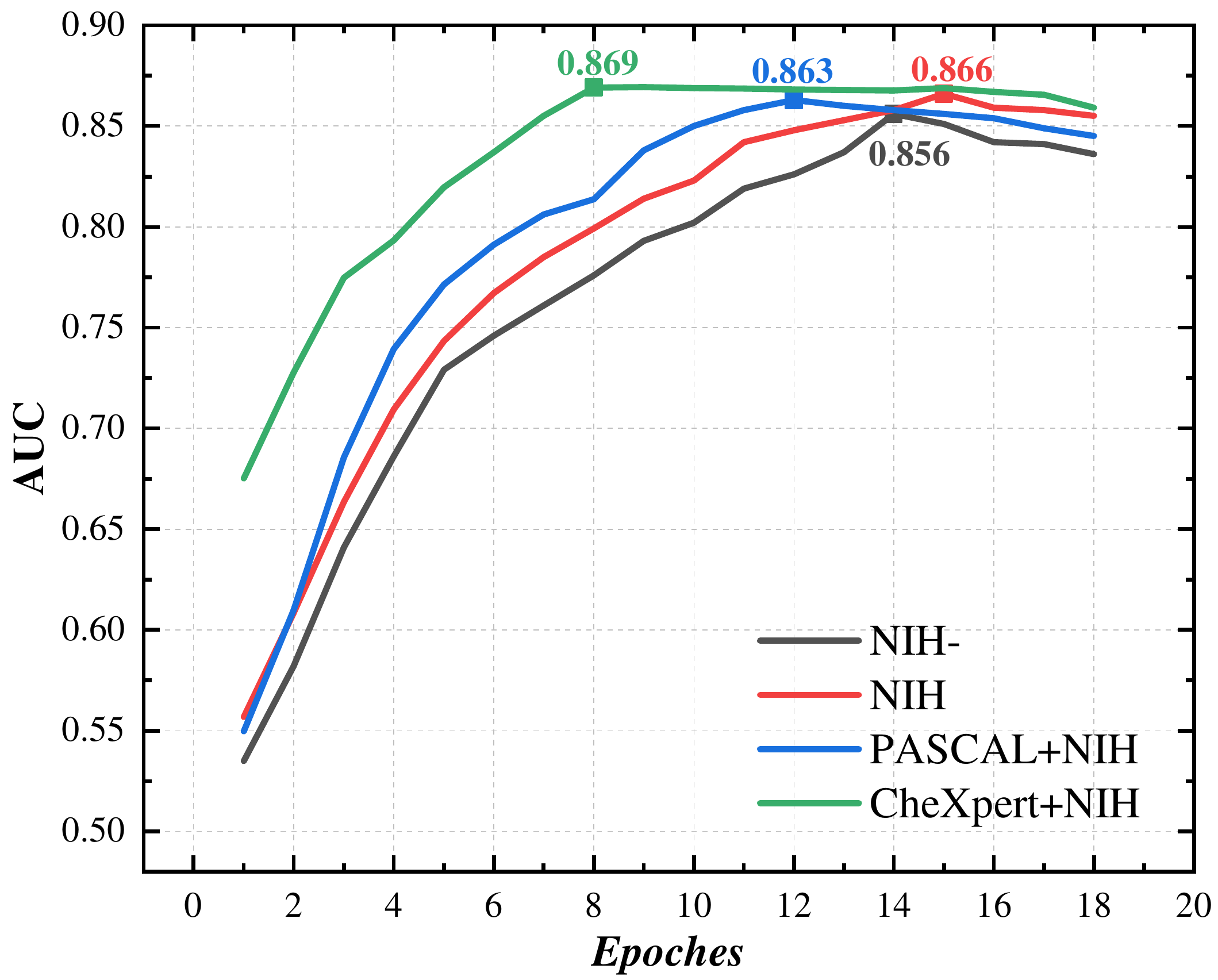}
\caption{Different experiment setting's mean AUC with experiment epoch.}
\label{domain}
\end{figure}

The related experimental results are shown in Fig.\ref{domain}.
``NIH'' means that we only use the NIH training data to initialize the parameters of agents; ``NIH-'' means that we remove the prior agents and only apply the main agent to make the final diagnosis. ``PASCAL+NIH'' means that we applied PASCAL data to train prior agents and NIH data to fine-tune all agents. ``CheXpert+NIH'' means that we applied CheXpert data to train prior agents. We can achieve the following observations:
\begin{itemize}
\item The prior agents can speed up the training of the main agent compared with that of ``NIH-''. ``NIH-'' achieved the worst performance. The other three methods introduced the prior agent. We also find that ``CheXpert+NIH'' has the fastest training speed, followed by ``PASCAL+NIH'', and ``NIH'' is the worst.  ``CheXpert+NIH'' and ``PASCAL+NIH'' applied the CHeXpert and Pascal data to intilize the parameters of prior agents. This means that the main agent uses more data, regardless of whether this data is consistent with the target domain data. It still allows the diagnostic agent to see more data. More data can provide better training results, and many deep learning algorithms have proved this conclusion. In these results, More data leads to faster convergence and better results. This also proves the necessity and effectiveness of the prior agents. 
\item Similar data with target domain can improve the final performance. We can find that ``CheXpert+NIH'' has the best results, followed by ``NIH'', and ``PASCAL+NIH'' is the worst. The reason rely on the quality of prior agent's parameter. CheXert data has a similar distribution to target domain data in NIH. PASCAL data has obviously different with target data. Thus, ``CheXpert+NIH'' has the best result. ``NIH'' has a better result than that of ``PASCAL+NIH''. The reason is that the large distribution difference with the target domain data has had a certain negative impact. This is also in line with people's common sense. After all, an orthopedic doctor needs a certain amount of study before he can become a chest doctor.
\end{itemize}

\subsection{The Performance of \texorpdfstring{$\epsilon$}{}-greedy}
\begin{figure}
\centering
\includegraphics[width=0.95\linewidth]{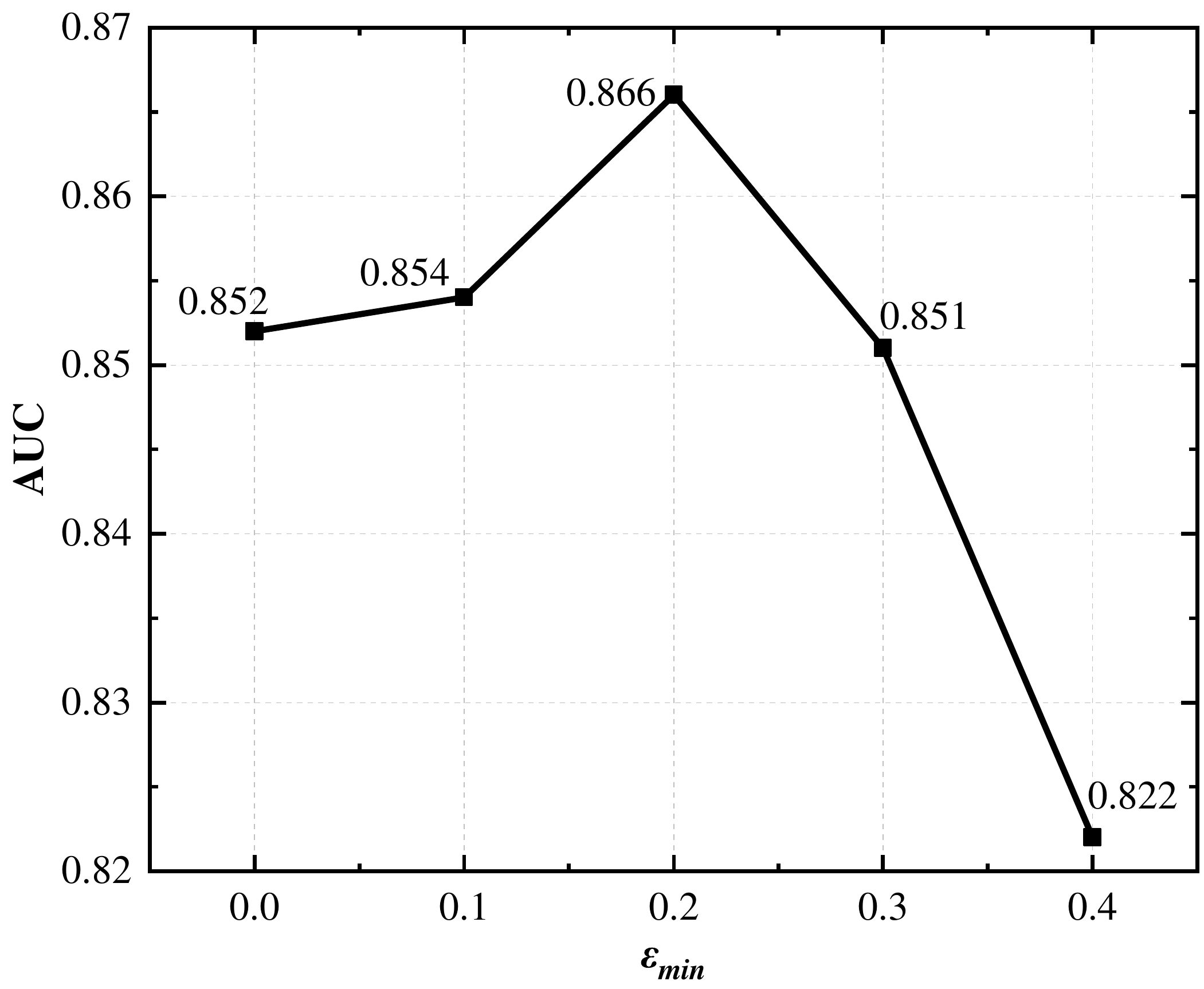}
\caption{The correlation of mean AUC and exploration.}
\label{expl}
\end{figure}

In the training step, we utilize the $\epsilon$-greedy algorithm to add some exploratory and speed up the convergence of the model. This is like a human being making some tentative decisions on certain issues without experience.
In this section, we adjust the value of the parameter $\epsilon$ to control the scale of exploration, and then observe its impact on the final result. The related experiment is shown in Fig.\ref{expl}.

The diagnostic result increases with the increase of $\epsilon$ value, and when $\epsilon$=0.2, the best result is achieved. But when $\epsilon$ continued to increase, the diagnostic results showed a rapid decline. When $\epsilon$ is equal to 0.2, it means that in one-fifth of the cases, the highest-scoring result was not chosen as the recognition result. If we continue to increase the value of $\epsilon$, it means that we discard the experience and increase the uncertainty of the diagnosis. This condition will lead to that the model introduces more wrong results in the training, which also leads to a rapid decrease in the results. From these results, we can find that appropriate exploration is effective and can speed up the convergence and training speed of the model. However, too much exploration will lose its original meaning, abandon the guiding significance of the real results, and introduce more uncertainties.
In general, this experiment demonstrated the effectiveness of training mechanism. This idea can be applied to many similar problems.

\subsection{Parameter Selection}
The parameters of transformer in diagnostic agents are worth further discussion. Here, we try to figure out how many decoder layers and multi-heads are in the inner structure of the Transformer in order to test its most important information-splitting and -combining functions in this vision task. At first, we set up 2 layers of transformer decoders, each with 4 heads, and got an average AUC of 0.866. Under the condition of 4 heads, we achieve a mean AUC of 0.843, 0.866, 0.853, 0.836, 0.829, and 0.825 when we set 1-6 layers, respectively. Under the condition of 2 layers, we achieve a mean AUC of 0.837, 0.849, 0.866, and 0.826 when we set 1, 2, 4, and 8 heads, respectively. 

\begin{figure}[t]
\centering
\includegraphics[width=0.95\linewidth]{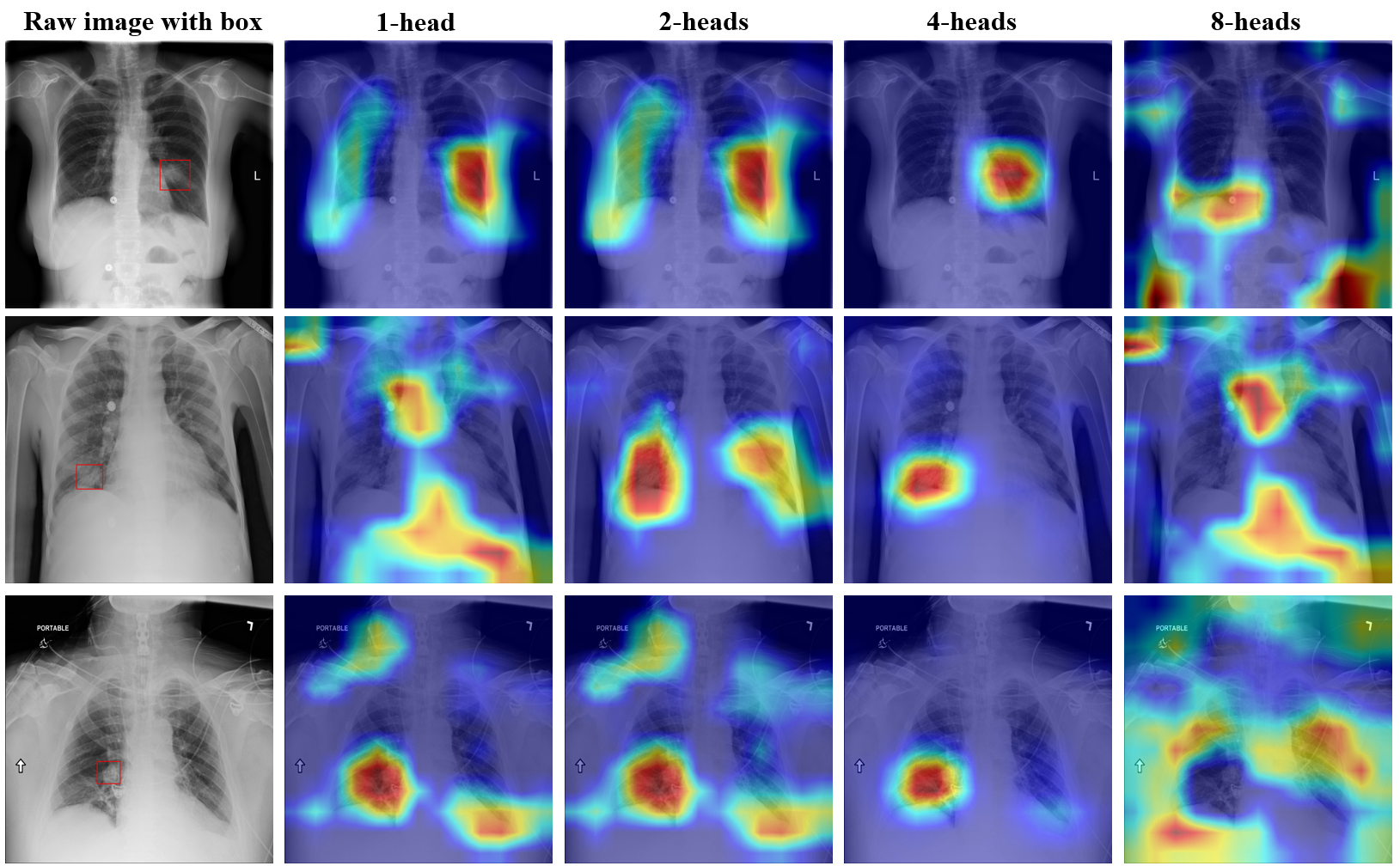}
\caption{Visualize results of average the maps of all heads. The leftmost column is the input CXR image, and disease areas are marked according to official documents by us, diseases from the top to bottom are ``Mass'', ``Pneumonia'', ``Mass'', the right 4 columns average the maps of all heads, 1, 2, 4, and 8 heads, respectively. Best view in colors.}
\label{attmaps}
\end{figure}

\begin{figure}
\centering
\includegraphics[width=0.95\linewidth]{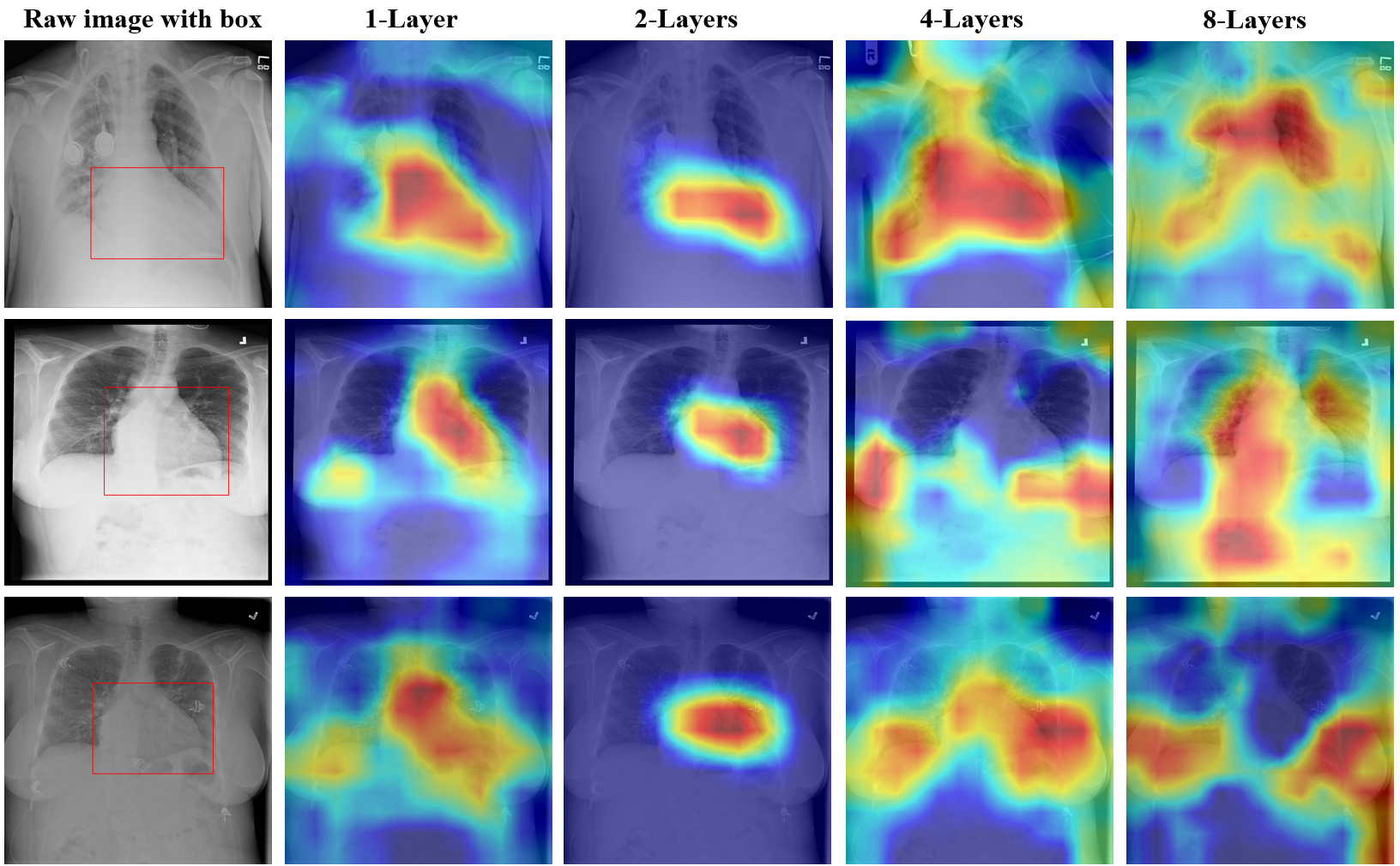}
\caption{Visualize results of layer attention maps, all pathologies are ``Cardiomegaly''. Best view in colors.}
\label{layeratt}
\end{figure}

To further verify the impact of layers and multi-headed transformers, we chose to visualize some attention maps of transformer structure. We are also interested in finding out the role of multi-head attention in this task. We plot the mean of each head’s cross-attention maps, which represent the similarities of a given query and extracted spatial features. 
From Fig.\ref{attmaps}, we can see that when the number of heads equals to 4, the attention maps better identify the lesion of thorax diseases than others, it testify in this specific task, it is good to use 4 heads. Besides, we can see that when the multi-heads are not enough, the attention weights are scattered and not very accurate, and when there are 8 multi-heads, the attention maps are chaotic, which means the multi-head attention mechanism introduces some redundant information and adds some interference to the classification task, which indicates that the redundant heads are not utilized as the other four heads already collect sufficient information for classification.
Similar to the layer attention map visualization experiment shown in Fig.\ref{layeratt}, we can find that when the number of layers of the transformer decoder equals 2, the visualization result shows better learned information than others, proving that the layer setting is important in this task.

\subsection{Algorithm Complexity}
The computational consumption is also a factor that should be considered for clinical usage. In general, our computational resource consumption is moderate compared to other methods.

We mostly focus on the model parameters and the size of the input image that affect how much GPU power our proposed MARL framework needs.
For a certain input CXR image resolution, e.g. 256 $\times$ 256, we use Resnet-18 and Resnet-50 in our framework, and the scale of model parameters in our proposed framework increases from 39.87M to 175.89M.

\begin{table*}[htbp]
  \centering
  \caption{Floating point operations with different backbone and resolution}
    \begin{tabular}{c|c|c|c}
    \cmidrule{1-4}
    Methods & Backbone & Resolution & FLOPs(GMACs) \\
    \cmidrule{1-4}
    Yao \textit{et al.}\citet{yao2018weakly} &DenseNet &224 &13.57\\
    Wang \textit{et al.}\citet{wang2017chestx} & AlexNet & 224 &14.54\\
    Wang \textit{et al.}\citet{wang2017chestx} & VGGNet & 224 &30.95\\
    LLAGNet \citet{chen2019lesion} & DenseNet &256 &34.96\\
    \hline
    Ours & ResNet & 256   & 63.57 \\
    Ours & VGGNet & 256   & 42.78 \\
    Ours & DenseNet & 256   & 52.31 \\
    Ours & ResNet & 512   & 159.96 \\
    \cmidrule{1-4}
    \end{tabular}%
  \label{flop}%
\end{table*}%

Taking a 256 $\times$ 256 input image as an example, as shown in Table.\ref{flop}, the FLOPs (floating point operations) for our proposed framework are 63.57G MACs (Multiplication and Accumulation) to 42.78G MACs from Resnet-50 to VGGNet, which the range of change is not small. In practice, training our proposed framework with 256 $\times$ 256 images and Resnet-18 or Resnet-50 on an NVIDIA GeForce RTX 3090 GPU with 24 GB memory costs nearly 0.033 or 0.145 seconds per image. Large-resolution images would cost much more time,
e.g., when we use 512 $\times$ 512 as the input CXR image resolution with Resnet-18 or Resnet-50 training on the same GPU, it costs nearly 0.069 or 0.251 seconds per CXR image. By the way, loading data into memory would take a lot of time compared to a smaller input resolution.
The FLOPs increase about 96.39G (63.57G vs. 159.96G) MACs for  512 $\times$ 512 with Resnet-18 as our backbone compared with input resolution 256 $\times$ 256, respectively.

Besides, the GPU consumption is extremely different when loading the training data with different input image sizes or batch sizes. In our experiment, when we adopt Resnet-18 as our backbone, training our proposed framework with 16 images (256 $\times$ 256) in a mini-batch costs about 2.4 GB of GPU memory. When the input is set to 32 images (256 $\times$ 256) in a mini-batch, 3.1 GB of GPU memory is required. While the input is set to 32 images (512 $\times$ 512) in a mini-batch, nearly 5.0 GB of GPU memory is required. In the above experiments, training with images of a higher resolution leads to better performance, but the amount of GPU memory used also goes up.

\section{Conclusion}
\label{sec:conclution}
In this paper, we propose a new multi-agent reinforcement learning framework to solve the multi-label CXR image classification problem. This framework uses diagnostic agents' previous knowledge to guide their learning, just like how a person learns. Prior knowledge is learned from the pre-trained model based on old data or similar data from other domains, which can effectively reduce the dependence on target domain data and speed up convergence. 2) The framework of reinforcement learning can make the diagnostic agent as exploratory as a human and improve the accuracy of diagnosis through continuous exploration. This design makes the whole model more intelligent and more in line with human learning rules. Meanwhile, the method can effectively solve the few-shot model learning problem and improve the model's generalization ability.
We evaluated our proposed method on two public data sets, NIH ChestX-ray14 and CheXpert. The experimental results demonstrate the performance of our approach. 

We next briefly discuss the limitations of our proposed method and future work possibilities. Firstly, our proposed MARL framework adopts the simplest form of multi-agent reinforcement learning, and more complicated multi-agent relationships are worth further study. Secondly, as we introduce reinforcement learning into the traditional classification framework, we will consume more computing resources and train much more slowly than in the traditional classification format, we will try to make our models lightweight in the future. We believe that further research will help us solve these issues. 

\section*{Acknowledgments}
This work was supported in part by the National Natural Science Foundation of China (62272337) and the Natural Science Foundation of Tianjin (16JCZDJC31100, 16JCZDJC31100).

\bibliographystyle{model2-names.bst}\biboptions{authoryear}
\bibliography{mybibliography}



\end{document}